\documentclass[twocolumn,fleqn]{article}
\usepackage{espcrc2}
\usepackage{graphics}
\begin{document}

\onecolumn{
\title{Bayesian Analysis of Many-Pole Fits of Hadron Propagators
 in Lattice QCD }
\author{D.~Makovoz\address{
{\em Department of
Physics, Ohio State University, Columbus, OH 43210,   USA}}}

\begin{abstract}
We use Bayes' probability theorem to analyze 
many-pole fits of hadron propagators. 
An alternative method of estimating values and 
uncertainties of the fit parameters is offered, which has certain
advantages over the conventional methods.
The probability distribution of the parameters of a fit is 
calculated.
The relative probability of various models is calculated.
\end{abstract}
\maketitle
}

\section{Introduction}
A common procedure in Lattice QCD is to calculate a correlation function in a
certain channel and then fit it as a sum of several exponentials
\cite{QCDPAX,UKQCD,dong,my}. 
The parameters of the fits are estimated by
 minimizing $\chi^2$. To find the errors,
ideally the calculations should be repeated many times,
but this is impractical.
Usually the jacknife  or the bootstrap is performed.
Instead here we use Bayes's theorem to derive the parameters' 
probability distribution for given data from the probability of the
data for given parameters.
Usually one determines the number of poles by 
comparing $\chi^2$ for different models.
It is very often ambiguous.
Here we use the parameters probability distribution
to calculate relative probabilities of the possible models with given data.
We perform the Bayesian analysis for some model 
functions imitating Lattice QCD propagators.
Then we apply this method to analyze SU(2) hadron propagators.

\section{Bayes' theorem and its applications}

$P(A|B)$ is the conditional probability that proposition $A$ is true,
given that proposition $B$ is true. Bayes' theorem reads:
\begin{eqnarray}
P(T|D,I) P(D|I) = P(D|T,I) P(T|I),
\end{eqnarray}
where $T$ is the theoretical model to be tested, 
$D$ is the data, and $I$ is the prior information. 
$P(T|D,I)$ is the posterior probability of the theoretical model.
$P(T|I)$ is the prior probability of the theoretical model.
$P(D|I)$ is the prior probability of the data; it will be always absorbed into
the normalization constant.
$P(D|T,I)$ is the direct probability of the data. 
For shortness $I$  will be implicit in all formulae henceforth. 

We are interested in calculating the posterior probability of a theoretical
 model and its parameters $\{c,E\}$:
\begin{eqnarray}
P(\{c,E\}|D)=\frac{P(\{c,E\})}{P(D)} P(D|\{c,E\}|). \nonumber
\end{eqnarray}

Given $P(\{c,E\})$, and $P(D|\{c,E\})$ we can~\cite{bret}: 

I. Calculate the posterior probability
density $P(E_n|D)$ 
[$P(c_n|D)$] for the parameters $E_n$ [$c_n$] :
\begin{eqnarray}
P(E_n|D) = \int 
\Pi \hspace{-0.05in}\raisebox{-1ex}{\scriptsize{$i$}}\hspace{0.05in} dc_i
\hspace{0.1in}\Pi \hspace{-0.2in}\raisebox{-1ex}{\scriptsize{$i\neq n$}}
dE_i\hspace{0.05in} P(\{c,E\}|D). \nonumber
\label{eq:distr}
\end{eqnarray}

II. Calculate the average values  $\overline{E}_i$ 
 and the standard deviations $\sigma_{E_i}$ (similarly for $c_i$) 
\begin{eqnarray}
\overline{E}_n = \int dE_n\hspace{0.05in} E_n P(E_n|D), \nonumber \\
\sigma_{E_i}^2 = \int dE_n\hspace{0.05in} (E_n-\overline{E_n})^2 P(E_n|D),
\label{eq:value}
\end{eqnarray}
provided $P(E_n|D)$ is normalized.

III. Compare several models $T_i$, 
(for example one pole, two pole and three pole models).
 We cannot find the absolute probability of a theory,
 since we do not have 
the ``complete set'' of theories. But we can calculate the relative
probabilities of two theories:
\begin{eqnarray}
\frac{P(T_i|D)}{P(T_j|D)} = \frac{P(T_i) P(D|T_i)}
{P(T_j) P(D|T_j)}.
\label{eq:tot_rat}
\end{eqnarray}
$P(D|T)$ can be obtained from $P(D|\{c,E\})$ by integrating over all 
parameters of the theory:
\begin{eqnarray}
P(D|T)= \int \{dc\} \{dE\} P(\{c,E\}) P(D|\{c,E\}). 
\label{eq:tot_pr}
\end{eqnarray}
For the prior  probability $P(\{c,E\})$ of the parameters of a model
in section \ref{sec:modelselection}
we make the ``least informative'' assumption~\cite{bret}
$P(c,E)$  $dc$ $dE$ $\sim$ $dc/c$ $dE/E$.
This form is scale invariant.
Priors $P(\{c,E\})$ should be normalized. 

The direct probability $P(D|\{c,E\})$ of the data $D$ can be calculated 
relatively easily if the data is Gaussian distributed.
We generate ``fake'' data   to be used in the analysis.
We use an n-pole model, add noise $e(t)$ to generate a sample of  
$N$ ``propagators'' $g_{\alpha}(t)$= $\sum_{i=1}^n$ $c_i e^{-E_i t}+e(t)$
 ($\alpha=1,..,N$),
calculate the average
 $G(t)$ and  estimate the covariance matrix $C(t,t')$ from the data.
Here t is the discrete, ``lattice'' time.
We vary the number of ``propagators'' to control 
the noise level  in the data.
Here we use $N=360$ and $N=3600$, which corresponds to a 
decrease in the noise level  
by a factor of 3. 

The probability distribution of $G(t)$ is \cite{TASI}
\begin{eqnarray}
P(D|\{c,E\})= e^{-\chi^2(\{c,E\})/2} ,
\label{eq:dir_pr_D}
\end{eqnarray}
where $\chi^2$ is calculated using the full covariance matrix \cite{TASI}.
The individual $g_\alpha(t)$ need not be Gaussian distributed,
 as long as we average
over enough of them  so that $G(t)$ are. 
Gaussian distribution of the ``fake'' data is ensured. 

\section{Estimating the Parameters.}

\subsection{1 pole data.}

We generate  data $D$ for the one pole model with 
$c_1^{in}=0.15$ and $E_1^{in}=0.485$.
We use the one pole model to fit the data. 
 Here we assume the prior probability of the data $P(\{c,E\})$ 
to be constant.
Then the posterior probability of the parameters $P(\{c,E\}|D)$
 is up to a constant equal to the direct probability of the data given
by equation  (\ref{eq:dir_pr_D}).
The posterior probability density for  $E_1$ 
\begin{eqnarray}
P(E_1|D) = \frac{\int dc_1 \hspace{0.05in} e^{-\chi^2(c_1,E_1)/2}}
{\int dc_1 dE_1 \hspace{0.05in} e^{-\chi^2(c_1,E_1)/2}}.
\label{eq:PcE}
\end{eqnarray}
It begs for the Monte Carlo integration with the Metropolis
algorithm. 

We generate a set of points $(c_1,E_1)$. Every point is characterized by 
$\chi^2(c_1,E_1)$. We sample the vicinity of the minimum
of $\chi^2(c_1,E_1)$ [maximum of $exp{(-\chi^2(c_1,E_1))}$](Fig.1).

%\begin{picture}(,150)
%\includegraphics{chis.ps}
%\end{picture}

\begin{picture}(,130)
\newsavebox{\onechi}
\newsavebox{\onechicap}
\savebox{\onechi}(50,80)
{\input{chi1.tex}}
\savebox{\onechicap}(50,30){\parbox{2.9in}
 {Figure 1. $\chi^2$ vs. the point number $n$.}}
\put(71,30){\usebox{\onechi}}
\put(71,5){\usebox{\onechicap}}
\end{picture}

We make a scatter plot $c_1$ vs.$ E_1$ (Fig.~2) to 
visualize this distribution. The density of the points
is proportional to the weight $exp{(-\chi^2(c_1,E_1))}$.

\begin{picture}(,160)
\newsavebox{\onepole}
\newsavebox{\onepolecap}
\savebox{\onepole}(50,100)
{\input{one_pole.tex}}
\savebox{\onepolecap}(50,30){\parbox{2.9in}
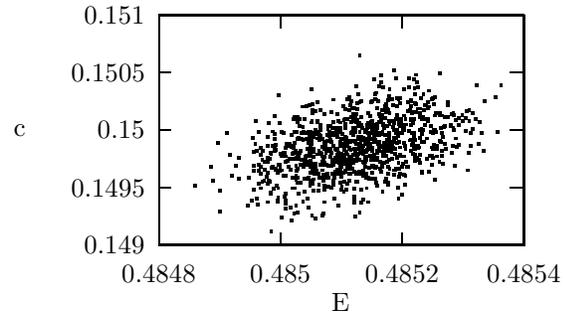
{Figure 2. Scatter plot of $\chi^2$ in ($c,E$) space.}}
\put(71,40){\usebox{\onepole}}
\put(71,0){\usebox{\onepolecap}}
\end{picture}

Taking integrals (\ref{eq:PcE}) is equivalent to making a histogram with 
steps big enough to make the distribution smooth (Fig.3).

\begin{picture}(,130)
\newsavebox{\onee}
\newsavebox{\oneecap}
\savebox{\onee}(50,90)
{% GNUPLOT: LaTeX picture
\setlength{\unitlength}{0.240900pt}
\ifx\plotpoint\undefined\newsavebox{\plotpoint}\fi
\begin{picture}(900,450)(0,0)
\font\gnuplot=cmr10 at 10pt
\gnuplot
\sbox{\plotpoint}{\rule[-0.200pt]{0.400pt}{0.400pt}}%
\put(219.0,134.0){\rule[-0.200pt]{4.818pt}{0.400pt}}
\put(197,134){\makebox(0,0)[r]{0}}
\put(816.0,134.0){\rule[-0.200pt]{4.818pt}{0.400pt}}
\put(219.0,224.0){\rule[-0.200pt]{4.818pt}{0.400pt}}
\put(197,224){\makebox(0,0)[r]{0.05}}
\put(816.0,224.0){\rule[-0.200pt]{4.818pt}{0.400pt}}
\put(219.0,315.0){\rule[-0.200pt]{4.818pt}{0.400pt}}
\put(197,315){\makebox(0,0)[r]{0.1}}
\put(816.0,315.0){\rule[-0.200pt]{4.818pt}{0.400pt}}
\put(219.0,405.0){\rule[-0.200pt]{4.818pt}{0.400pt}}
\put(197,405){\makebox(0,0)[r]{0.15}}
\put(816.0,405.0){\rule[-0.200pt]{4.818pt}{0.400pt}}
\put(219.0,134.0){\rule[-0.200pt]{0.400pt}{4.818pt}}
\put(219,89){\makebox(0,0){0.4848}}
\put(219.0,385.0){\rule[-0.200pt]{0.400pt}{4.818pt}}
\put(425.0,134.0){\rule[-0.200pt]{0.400pt}{4.818pt}}
\put(425,89){\makebox(0,0){0.485}}
\put(425.0,385.0){\rule[-0.200pt]{0.400pt}{4.818pt}}
\put(630.0,134.0){\rule[-0.200pt]{0.400pt}{4.818pt}}
\put(630,89){\makebox(0,0){0.4852}}
\put(630.0,385.0){\rule[-0.200pt]{0.400pt}{4.818pt}}
\put(836.0,134.0){\rule[-0.200pt]{0.400pt}{4.818pt}}
\put(836,89){\makebox(0,0){0.4854}}
\put(836.0,385.0){\rule[-0.200pt]{0.400pt}{4.818pt}}
\put(219.0,134.0){\rule[-0.200pt]{148.635pt}{0.400pt}}
\put(836.0,134.0){\rule[-0.200pt]{0.400pt}{65.284pt}}
\put(219.0,405.0){\rule[-0.200pt]{148.635pt}{0.400pt}}
\put(45,269){\makebox(0,0){P(E)}}
\put(527,44){\makebox(0,0){E}}
\put(219.0,134.0){\rule[-0.200pt]{0.400pt}{65.284pt}}
\put(248,135){\rule{1pt}{1pt}}
\put(279,137){\rule{1pt}{1pt}}
\put(310,141){\rule{1pt}{1pt}}
\put(341,153){\rule{1pt}{1pt}}
\put(372,189){\rule{1pt}{1pt}}
\put(403,221){\rule{1pt}{1pt}}
\put(433,262){\rule{1pt}{1pt}}
\put(464,310){\rule{1pt}{1pt}}
\put(495,355){\rule{1pt}{1pt}}
\put(526,392){\rule{1pt}{1pt}}
\put(557,369){\rule{1pt}{1pt}}
\put(588,351){\rule{1pt}{1pt}}
\put(619,283){\rule{1pt}{1pt}}
\put(649,244){\rule{1pt}{1pt}}
\put(680,204){\rule{1pt}{1pt}}
\put(711,172){\rule{1pt}{1pt}}
\put(742,156){\rule{1pt}{1pt}}
\put(773,139){\rule{1pt}{1pt}}
\put(804,136){\rule{1pt}{1pt}}
\put(834,135){\rule{1pt}{1pt}}
\end{picture}}
\savebox{\oneecap}(50,30){\parbox{2.9in}
 {Figure 3. Probability distribution $P(E)$. }}
\put(71,30){\usebox{\onee}}
\put(71,0){\usebox{\oneecap}}
\end{picture}

Both equation (\ref{eq:value}) and jacknife give 
the same values for the parameters and 
errors of $E= 0.4851(1)$ and $c=0.1499(2)$. 
 The present approach conserves  computational time
compared to the jacknife. 
If one uses simulated annealing to fit the data, one  covers the
same regions in the ($c_i$,$E_i$) space 
as needed to calculate  probability distributions~(\ref{eq:distr}). 
With the probability distributions one immediately obtains the 
parameters and errors, 
whereas with the jacknife the fitting has to be repeated $N$ times.

\subsection{2 Pole data}
We repeat the analysis  performed in section 3.1  
for two-pole data  when the poles are well separated:
 $c_1^{in}$=0.1, $c_2^{in}$=0.1 $E_1^{in}$=0.5 
$E_2^{in}$=0.6.
We use the two-pole model to fit the data. 
The probability distributions for the energies and
coefficients are obtained and the parameters are estimated 
just as in section 3.1. 

The only complication is that now we have to deal with
the 4-d space of parameters $(c_1,E_1,c_2,E_2)$. 
We perform a Monte Carlo integration as described above.
The 4-d probability distribution is visualized by projecting it onto two planes
$(c_1,E_1)$ and $(c_2,E_2)$.
Each blob in Fig. 4 is a projection of the 4-d distribution on a 2-d plane.
Each blob represents the probability distribution for one pair of $(c_i,E_i)$ 
after the second pair has been integrated out.

\begin{picture}(,150)
\newsavebox{\twopole}
\newsavebox{\twopolecap}
\savebox{\twopole}(50,100)
{\input{two_pole.tex}}
\savebox{\twopolecap}(50,40){\parbox{2.9in}
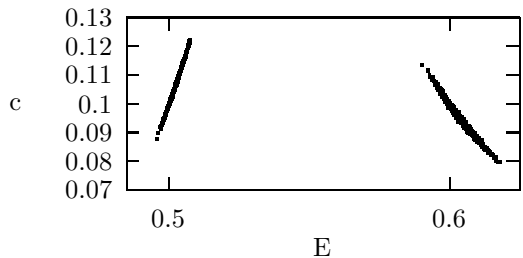
 {Figure 4. Scatter plot of $\chi^2$ 
 projected on $(c_1,E_1)$ and $(c_2,E_2)$  planes and
combined on one plane.}}
\put(71,40){\usebox{\twopole}}
\put(71,0){\usebox{\twopolecap}}
\end{picture}

%\begin{figure}
%\input{/net/campbell/04/makovoz/FIT/figures/two_pole.tex}
%\caption{Scatter(density) plot of $chi^2$ in the space of its parameters
% $E_1$,$E_2$,$c_1$, and $c_2$ projected on the two respective planes and
%combined on one plane $c,E)$.} 
%\label{fig:twopole}
%\end{figure}

\section{Model selection}
\label{sec:modelselection}

Unless one has some prior knowledge, determining 
the number of poles present in the data  
based on comparing $\chi^2$  for models with different numbers of poles
 is very often ambiguous.

Table~\ref{tab:chi} contains the values of $\chi^2$ obtained when 1 and 2 pole
data are fit with 1 and 2 pole models. 
We use the one pole data from section 3.1 and generate 
2 pole data with the poles close to each other:
$c_1^{in}=0.05$ $c_2^{in}=0.1$ $E_1^{in}=0.49$ $E_2^{in}=0.5$.
With one exception
the corresponding values of $\chi^2$  are close 
 and do not allow for a definite
answer.
\begin{table}[h]
\begin{tabular}{cccc}\hline
 data & \# of propagators  & model &$\chi^2/dof$  \\ \hline
 &  &1 pole & 0.61  \\ \cline{3-4}
 & 360 &  2 pole &0.57  \\ \cline {2-4}
1 pole & &  1 pole  & 1.2  \\ \cline{3-4}
  & 3600 &  2 pole & 1.3     \\ \hline \hline

 &  &1 pole &0.87  \\ \cline{3-4}
 & 360 & 2 pole &0.99   \\ \cline {2-4}
 2 pole & &   1 pole  & 2.0  \\ \cline{3-4}
 & 3600 & 2 pole& 1.2     \\ \hline
\end{tabular}
\caption{Values of $\chi^2$ for fitting 1 and 2 pole data 
of different noise levels with 1 and 2 pole  models.}
\label{tab:chi} 
\end{table}

We need to calculate the total probabilities ratio (\ref{eq:tot_rat}) 
to determine the number of poles in the data.
We assume $P(2 pole)=P(1 pole)$, i.e. {\em a priori} these two models
are equally probable.  
To calculate $ P(D|1 pole)$ 
 we substitute 
  (\ref{eq:dir_pr_D}) into equation (\ref{eq:tot_pr})
 (and similarly for $P(D|2 pole)$)
\begin{eqnarray}
P(D|1 pole)= \int dc_1 dE_1 \hspace{0.05in}\frac{ e^{-\chi^2(c_1,E_1)}}
{c_1 E_1} 
\label{eq:tot_pr1}
\end{eqnarray}
The integration is tricky since we are dealing  with a function that varies 
 rapidly in the multidimensional space.
We use  scatter plots to determine the areas of integration.
Table \ref{tab:ratio} contains integration results for the total probability
ratio $R=\frac{P(1 pole|D)}{P(2 pole|D)}$.
With a sufficiently low noise level the total probabilities ratio 
picks the correct model.
\begin{table}[h]
\begin{tabular}{ccc}\hline
data & \# of propagators &  R \\ \hline
 & 360  & 3 \\ \cline{2-3}
1 pole & 3600 & 30 \\ \hline
 & 360 & 3 \\ \cline{2-3}
2 pole & 3600& 0.02 \\ \hline
\end{tabular}
\caption{Total probabilities ratio $R$.}
\vspace{-0.3in}
\label{tab:ratio}
\end{table}
If we estimate the parameters 
of the 2 pole data with 3600 propagators 
using equation (\ref{eq:value}), we get  the input
parameters back within the error bars.
The jacknife here gives unreasonably large errors 
 because  
$\chi^2$ has several minima in the ($c_1,c_2,E_1,E_2$) space.
It can be seen from the graph of $\chi^2$ vs. $n$ (Fig.7),
and they can be also clearly identified on the scatter plot (Fig.8).
Two minima ( 1 and 3)  have the same $\chi^2$.  
When we perform the jacknife, the blind fitting finds either minimum 1 or 3,
which results in the unreasonably large error estimates.

\begin{picture}(,165)
\newsavebox{\twopolonechi}
\newsavebox{\twopolonechicap}
\savebox{\twopolonechi}(50,100)
{\input{two_pol_one_chi.tex}}
\savebox{\twopolonechicap}(50,30){\parbox{2.9in}
 {Figure 7. $\chi^2$ has multiple minima.}}
\put(71,50){\usebox{\twopolonechi}}
\put(71,20){\usebox{\twopolonechicap}}
\end{picture}

\begin{picture}(,135)
\newsavebox{\twopolone}
\newsavebox{\twopolonecap}
\savebox{\twopolone}(50,100)
{\input{two_pol_one.tex}}
\savebox{\twopolonecap}(50,30){\parbox{2.9in}
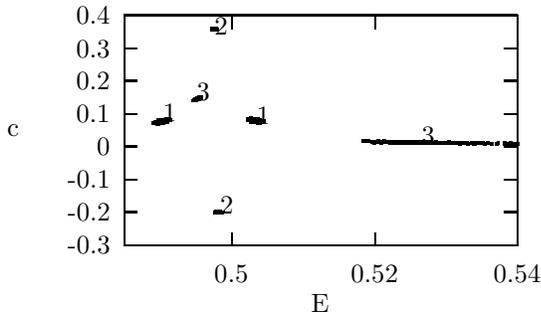
 {Figure 8. Three minima of $\chi^2$  from Fig.7
projected on the two-dimensional 
plane $(c,E)$.}}
\put(71,40){\usebox{\twopolone}}
\put(71,0){\usebox{\twopolonecap}}
\end{picture}

\section{Analysis of SU(2) data}
Here we analyze hadron propagators in the pseudoscalar channel. The detailed 
description of these data is given in \cite{my,colormy}.
The coupling constant $\beta=2.5$,
lattice spacing $a=0.09\pm0.012$ fm,
the lattice size is $12^3\times 24$. 
The propagators analyzed here were calculated with $\kappa=0.146$ 
for 360 configurations.
The point source is at $t=5$. The fit is performed in the time range 6-20.

We repeat the analysis for 60 and 360 configurations to 
study the data with different noise levels. For 60 configurations 
the $\chi^2/dof$ is 1.0 for the 3 pole fit
 and 1.17 for  the 4 pole fit, 
and the total probabilities ratio is 
 $\frac{P(3 pole|D)}{P(4 pole|D)}\sim 1$. For 360 configurations
the $\chi^2/dof$ is 0.52 for
 the 3 pole model fit and 0.69 for
  the 4 pole model fit, and
the total probabilities ratio for 360 configurations is
 $\frac{P(3 pole|D)}{P(4 pole|D)}\sim 10$.
Here again, for  low noise data
we are able  to choose between 
two models based on a qualitative estimate given by 
 the total probabilities ratio.

\section{Conclusion}

A new method has been  introduced that can be used to analyze many-pole fits
of hadron propagators in Lattice QCD. 
It has been used to estimate the many-pole model parameters  and their 
uncertainties.  
It works in the presence of multiple minima when the jacknife (at least in its 
simpleminded form) fails. It cuts the computational time.
The new method has been used to calculate 
relative probabilities for different models, which can be crucial 
in making the optimal choice of a model. 
The scatter plots, which have been introduced as an auxiliary tool for the
multidimensional integration, can be used as an independent tool 
for the many pole fit analysis. 

I wish to thank  Greg Kilcup for his suggestion to use  Bayesian analysis.

\end{document}